\documentclass[10pt,letterpaper]{article}
\usepackage[top=0.85in,left=2.75in,footskip=0.75in,marginparwidth=2in]{geometry}

\usepackage[utf8]{inputenc}

\usepackage{cite}

\usepackage{nameref,hyperref}
\usepackage[numbers]{natbib}

\usepackage{microtype}
\DisableLigatures[f]{encoding = *, family = * }

\raggedright
\usepackage{changepage}

\usepackage[aboveskip=1pt,labelfont=bf,labelsep=period,singlelinecheck=off]{caption}

\makeatletter
\renewcommand{\@biblabel}[1]{\quad#1.}
\makeatother

\usepackage{lastpage,fancyhdr,graphicx}
\usepackage{epstopdf}
\fancyheadoffset[L]{2.25in}
\fancyfootoffset[L]{2.25in}

\usepackage{color}

\definecolor{Gray}{gray}{.25}

\usepackage{graphicx}

\usepackage{sidecap}

\usepackage{wrapfig}
\usepackage[pscoord]{eso-pic}
\usepackage[fulladjust]{marginnote}
\reversemarginpar
\usepackage{times}
\usepackage{wrapfig, frame}
\usepackage{multirow}

\begin{document}
\vspace*{0.35in}

\begin{flushleft}
{\Large
\textbf\newline{Interplay between physics self-efficacy, calculus transfer ability, and gender}
}
\newline
\\
Christopher J. Fischer\textsuperscript{*},
Jennifer Delgado,
Sarah LeGresley
\\
\bigskip
Department of Physics and Astronomy, University of Kansas, 1251 Wescoe Hall Drive, Lawrence, KS, 66049
\\
\bigskip
* shark@ku.edu

\end{flushleft}

\section*{Abstract}
    We present our initial work to develop an assessment of calculus proficiency in the context of introductory physics (i.e., calculus transfer to physics), including a comparison of calculus transfer ability with physics self-efficacy and how these attributes intersect with student gender.  Although students demonstrated an improvement in their calculus transfer ability, most nevertheless displayed a decrease in their physics self-efficacy.  Similarly, while women consistently exhibited lower physics self-efficacy than men, women demonstrated larger gains in their proficiency with calculus transfer to physics.  We also discuss future work to better evaluate the interdependence of mathematics self-efficacy, physics self-efficacy, and ability to transfer calculus to physics.


\section{Introduction}

Nearly all undergraduate STEM degree programs require courses in introductory physics since the content of these courses is a foundation to those disciplines. The emphasis on applied mathematics in these courses is believed to be responsible for the observed positive correlation between mathematics ability and student performance in these courses \cite{hudson1977correlation,cohen1978cognitive,hudson1981correlation,brekke1994some} and the motivation for some of these courses requiring prerequisite classes in mathematics.  While this prerequisite structure implicitly assumes that students are able to transfer skills and knowledge from their mathematics courses to their science and engineering courses, the results of several studies indicate that such transference is difficult \cite{britton2002students,nguyen2003initial,Ozimek2005,cui2006assessing,Rebello2007,new2012researching,Mikula2014}.  Although methods of improving calculus proficiency and possibly transfer in introductory physics have been suggested \cite{PhysRevPhysEducRes.15.020126,kezerashvili2007transfer,wagner2012representation,turcsucu2017teachers}, the lack of a robust instrument for assessing calculus skills in a physics context hinders the further refinement of these and other improvements in the curriculum of introductory physics that target improvements in calculus transfer. 

We report here on our initial efforts to develop an assessment of student proficiency with integral calculus in the context of introductory physics, including analysis of the results obtained with this assessment in the second semester of calculus-based introductory physics during Fall 2020.  We independently assessed physics self-efficacy in this same course using an established instrument \cite{marshman2018female}.  This allowed us to investigate whether any correlations exist between calculus transfer ability and physics self-efficacy or, more broadly, between calculus proficiency and physics self-efficacy.  While a correlation between calculus transfer proficiency and physics self-efficacy cannot yet be resolved by the data gathered so far, these preliminary results do suggest a potential influence of student gender on calculus transfer ability.  We also discuss future work to refine these assessments and associated analysis methodologies.

\section{Background}

Students pursuing calculus-based STEM degree programs (e.g., physical sciences and engineering) at the University of Kansas complete the first semester of calculus during their first semester.  General Physics I (the first semester of calculus-based introductory physics covering classical mechanics and thermodynamics) is taken during their second semester concomitant with the second semester of calculus.  General Physics II (the second semester of calculus-based introductory physics covering electricity and magnetism) is taken during their third semester at concomitant with the third semester of calculus.  Scheduling the calculus courses to be completed before or at the same time as the physics courses facilitates tasking students to use calculus continually in these physics courses \cite{PhysRevPhysEducRes.15.020126,Fischer2019a,Fischer2019b}.  This has also highlighted the need to improve the ability of our students to transfer their calculus skills to solving physics problems.

\subsection{Mathematics Transference}

We use the model proposed by Rebello \textit{et al.} \cite{Rebello2007}, which builds upon earlier work by Gagn\'{e} \cite{gagne308conditions} and Reddish \cite{redish2004theoretical}, to characterize knowledge transfer.  According to this model, transfer can be viewed as the creation of associations by the learner between prior knowledge and information or knowledge in a new problem presented to the learner.  This process of creating associations is further divided into \textit{horizontal transfer}, which involves assigning information in the new problem to an element of prior knowledge, and \textit{vertical transfer}, which involves connecting new knowledge in the new problem with an element of prior knowledge \cite{cui2006assessing,Rebello2007}.  In other words, horizontal transfer describes the use of new information in existing mental models, such as assigning numerical values given in a problem to the correct variables in an equation \cite{cui2006assessing}.  In contrast, vertical transfer describes the situation where the learner does not have an existing mental model that aligns with the new problem's information.  Rather, the learner recognizes elements of the new problem and constructs a new mental model, likely based on existing mental models \cite{cui2006assessing}.

Physics problems that task students with substituting numbers into given equations require only horizontal transfer and focus the attention of students on manipulating equations, rather than on thinking about the applicability of those equations or the underlying physics concepts they mathematically describe.  Such problems do not involve vertical transfer of knowledge as their solutions do not require the students to develop new mental models.  In contrast, many ``real-world'' problems, including those often encountered by students in upper-level science and engineering courses require vertical transfer, including vertical transfer of mathematics skills.  Indeed, students in these courses frequently encounter problems that do not involve a single identifiable equation or approach that should be used, and thus are tasked with developing new mental models, deciding between existing mental models, or combining multiple mental models together.  Therefore, incorporating new curriculum elements in introductory physics that target improving student ability with vertical transfer of mathematics would result in these courses preparing students better for the rest of their degree programs and their future careers.  Furthermore, since physics is often the first course taken by students in their degree programs that relies heavily on applied calculus, introductory physics courses provide an excellent venue for helping students with their calculus transfer skills.  

The process of revising the curriculum of introductory physics to help students improve their calculus transfer skills can be made more efficient through the use of a formative assessment of the vertical transfer of calculus to physics, which could quantify how curriculum changes affect calculus transfer proficiency.  The results reported here are our initial work to develop that assessment tool.

\subsection{Physics Self-Efficacy}

An individual's self-efficacy is a measure of their belief to succeed in a particular discipline or at a particular task \cite{sawtelle2012exploring,nissen2016gender,espinosa2019reducing}.  Self-efficacy can impact many aspects of learning including motivation \cite{prat2010interplay}, interest \cite{zimmerman2000self}, engagement \cite{schunk2002development}, and whether difficult tasks are perceived as challenges or threats \cite{watt2006role}.  For example, students with low self-efficacy are less likely to employ effective learning strategies than students with high self-efficacy \cite{zimmerman2000self,pintrich2003motivational}.  Self-efficacy has a strong influence over student performance and retention in STEM courses \cite{sawtelle2012exploring} and, unfortunately, several studies have shown that women tend to have lower self-efficacy in physics than men despite having similar or better performance in their physics courses \cite{nissen2016gender,marshman2018female,espinosa2019reducing}.  Of particular interest to our goal of developing a formative assessment of calculus transfer to physics is identifying the extent to which the gender imbalances for physics self-efficacy \cite{nissen2016gender,marshman2018female,espinosa2019reducing} and mathematics self-efficacy \cite{correll2001gender,hill2010so} influence calculus transfer proficiency.  Should a correlation between physics self-efficacy and calculus transfer ability exist, it may be possible to design new curriculum elements that target improvements in both simultaneously.

\section{Results}

Assessments were administered to students in General Physics II during the Fall 2020 semester.  The total enrollment in the course was 287 students, consisting of 212 males and 75 females.  Please note that we are relying upon institutional data for creating these cohorts of students.  These divisions do not therefore necessarily reflect how students identify personally.  The average ACT math scores for these two cohorts of students were $29.7 \pm 0.3$ (men) and $28.9 \pm 0.5$ (women).

The number of students completing each assessment is indicated below.  Not all students completed each assessment for a variety of logistical and other reasons.  Also, note that more students completed the physics self-efficacy assessment than the calculus transfer assessment.  This may have occurred since the physics self-efficacy assessment consisted of multiple choice questions and thus could be completed more easily and quickly than the calculus transfer assessment which had free response questions and thus required more effort from the students to complete.

\subsection{Assessing Calculus Transference}

General Physics II is a calculus-based introduction to electricity and magnetism.  As such, integral calculus constitutes the majority of the applied mathematics tasked to the students in the course, including calculating the electric and magnetic fields and potentials of distributions of electric charge, electric and magnetic flux, and the electric and magnetic forces and potential energies associated with the interactions of charged systems.  We therefore decided to structure our prototype calculus transfer assessment for General Physics II around integral calculus.  This initial assessment consists of the following three questions:

\vspace{2mm}

\noindent \textbf{Question 1.} If $x = y^2$, complete the following integral: $\int \limits_0^{L^2} y \hspace{1mm} dx$

\vspace{1.5mm}

\begin{wrapfigure}{r}{0.2\textwidth}
\label{fig:prelimmta1}
    \includegraphics[width=0.2\textwidth]{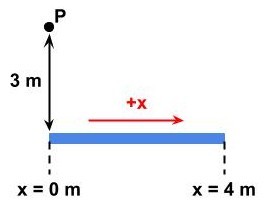}
      \label{fig:prelimmta}    
\end{wrapfigure}

\noindent \textbf{Question 2.} The electric potential $\left( V_e \right)$ a distance $r$ away from a point particle with electric charge $q$ is described by the following equation: $V_e = \frac{q}{4 \pi \epsilon r}$.  The straight and rigid rod shown in the figure to the right has a length of 4 m and a non-uniform linear electric charge density described by the following equation: $\left( 2 \pi \times 10^{-9} \mathrm{\frac{C}{m^2}}   \right) x$.  The variable $x$ in this equation denotes position along the $x$-axis as defined in the figure.  Determine the electric potential of this rod at the position $P$ that is a distance 3 m above one end of the rod as shown in the figure.  The variable $\epsilon = 1 \times 10^{-11} \mathrm{\frac{C}{Vm}}$.

\vspace{1.5mm}

\begin{wrapfigure}{r}{0.2\textwidth}
\label{fig:prelimmta2}
  \begin{center}
    \includegraphics[width=0.2\textwidth]{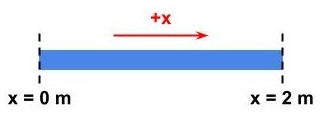}
  \end{center}
\end{wrapfigure}

\noindent \textbf{Question 3.}  The moment of inertia $\left( I \right)$ of a point particle with mass $m$ located a distance $x$ away from an axis of rotation is $I = mx^2$.  The straight and rigid rod shown in the figure to the right has a length of 2 m and non-uniform mass density described by the following equation: $\left( 5 \mathrm{\frac{kg}{m^3}}  \right) x^2$.  The variable $x$ in this equation denotes position along the $x$-axis as defined in the figure.  Determine the moment of inertia of this rod for rotations around an axis located at $x = 0 \mathrm{m}$.  

\vspace{2mm}

\noindent While question 1 is a pure calculus question, both question 2 and question 3 require transfer of calculus to physics; furthermore, one approach to solving the integral in question 1 is similar to the approach taught to students for solving integrals encountered in question 2 and question 3.  Students in General Physics II were tasked with determining the electric potential of a uniformly charged rod and thus were familiar with the underlying mathematical processes (\textit{i.e.}, the problem solving strategy) necessary to solve question 2.  Students did not explicitly solve problems involving non-uniformly charged objects during the semester, however, so that element of the problem is new.  Question 2 thus has elements of both horizontal and vertical transfer.  Since students at our institution have not used calculus to determine moments of inertia in General Physics I or General Physics II, solving question 3 requires vertical transfer.  However, by prompting students with the equation for the moment of inertia of a point particle, we hoped that student would recognize that the integration necessary for the solution is similar to the integration necessary to solve question 2.  

\begin{figure*}
  \includegraphics[width=1.0\textwidth]{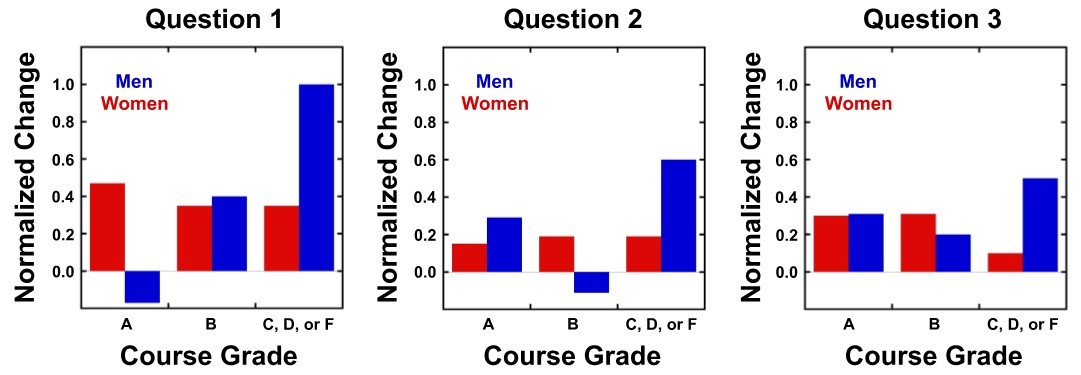}
  \caption{Normalized change \cite{marx2007normalized} for answering each question on the calculus transfer assessment correctly fractionated by course grade and student gender.  117 students (83 men and 34 women) earned an A, 61 students (50 men and 11 women) earned an B, and 37 students (32 men and 5 women) earned a C, D, or F. \label{fig:MTA_combined}}
\end{figure*}

We administered this calculus transfer assessment at the beginning (pre-test) and at the end (post-test) of Fall 2020 semester of General Physics II.  We administered the assessment online in a free response format and student responses were scored as correct or incorrect; responses that contained the correct integral were scored as correct, even if the subsequent integration was performed incorrectly.  The results of these assessments are shown in Table \ref{tab:MTAresults}.  The Pearson correlation coefficients for student responses to the questions on the calculus transfer assessment for the post-test are shown in Table \ref{tab:correlation_post}.  Finally, Figure \ref{fig:MTA_combined} shows these results fractionated by course grade and student gender.  

\begin{table}[htbp]
  \caption{Percentage of men (N = 165) and women (N = 50) answering questions on the mathematics transference assessment correctly.\label{tab:MTAresults}}
    \begin{tabular}{c c c c c}
         & &  \textbf{Pre-Test} & \textbf{Post-Test} & \textbf{Normalized Change \cite{marx2007normalized}} \\
         \hline
         \hline
         \multirow{2}{*}{\textbf{Question 1}} & Men & 56.4$\%$ &  73.9$\%$ & 0.40\\
         & Women & 64.0$\%$ &  66.0$\%$ & 0.06\\
         \multirow{2}{*}{\textbf{Question 2}} & Men & 4.8$\%$ &  21.2$\%$ & 0.17 \\
         & Women & 4.0$\%$ &  28.0$\%$ & 0.25 \\
         \multirow{2}{*}{\textbf{Question 3}} & Men & 14.5$\%$ &  37.0$\%$ & 0.26\\
         & Women & 8.0$\%$ &  36.0$\%$ & 0.30\\
    \end{tabular}
\end{table}

\begin{table}[htbp]
  \caption{Correlation between responses to questions on mathematics transfer post-test by all students.\label{tab:correlation_post}}
    \begin{tabular}{c c c c}
         & \textbf{Question 1} & \textbf{Question 2} & \textbf{Question 3}  \\
         \hline
         \hline
         \textbf{Question 1} & 1.000 & & \\
         \textbf{Question 2} & 0.064 & 1.000 & \\
         \textbf{Question 3} & 0.106 & 0.507 & 1.000\\
    \end{tabular}
\end{table}

\subsection{Assessing Physics Self-Efficacy}

\begin{figure*}
  \includegraphics[width=0.9\textwidth]{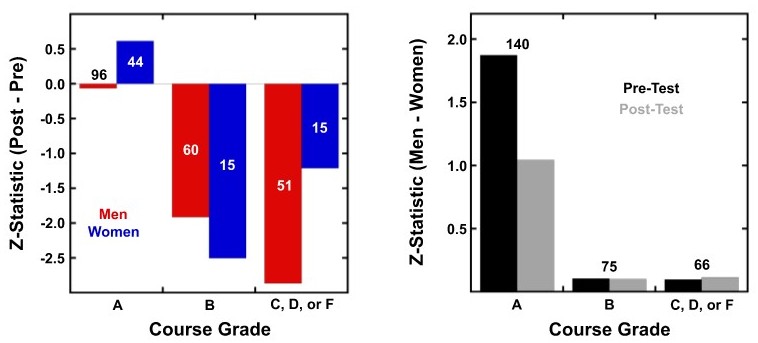}
  \caption{Normalized change for each question on the physics self-efficacy assessment fractionated by course grade and student gender.  While women earning an A in the course saw an increase in their self-efficacy during the course, the self-efficacy of all other students decreased.  This results in a decrease in the difference between the self-efficacy of men and women earning and A.  Men have a higher self-efficacy than women within each grade cohort, but the difference is most stark for students earning an A.  The number of students in each cohort is also shown. \label{fig:PSE_combined}}
\end{figure*}

We also assessed the physics self-efficacy of students in General Physics II during the Fall 2020 semester using the instrument previously developed by Marshman \textit{et al.} \cite{marshman2018female}.  This multiple choice survey consists of six questions, each of which involved 4-point Likert scales, with higher scores indicating higher levels of self-efficacy.  A single self-efficacy score for each student is determined by averaging the scores for the six questions.  Students completed the assessment at the beginning of the semester (pre-test) and again at the end of the semester (post-test).  The self-efficacy scores for students on these assessments are shown in Table \ref{tab:PSE} and Table \ref{tab:PSE_2}; the Z-statistics comparing the post-test and pre-test results and the Z-statistics comparing men and women are also shown in Table \ref{tab:PSE}.  Finally, Figure \ref{fig:PSE_combined} shows the Z-statistics from Table \ref{tab:PSE} fractionated by course grade and student gender.  

\begin{table}[htbp]
  \caption{Changes in Physics Self-Efficacy for men (N = 207) and women (N = 74) between the beginning and end of the semester. \label{tab:PSE}}
    \begin{tabular}{c c c c}
         &  \multirow{2}{*}{\textbf{Pre-Test}} & \multirow{2}{*}{\textbf{Post-Test}} & \textbf{Z-Statistic} \\
         & & & (Post - Pre) \\
         \hline
         \hline
         \textbf{Men}  & $2.81 \pm 0.02$ &  $2.71 \pm 0.03$ & -2.47\\
         \textbf{Women}  & $2.72 \pm 0.04$ &  $2.65 \pm 0.04$ & -1.18 \\
         \textbf{Z-statistic} & \multirow{2}{*}{1.92} & \multirow{2}{*}{1.20} \\
         (Men - Women) \\
    \end{tabular}
\end{table}

\begin{table}[htbp]
  \caption{Physics Self-Efficacy scores for men and women fractionated by course grade. \label{tab:PSE_2}}
    \begin{tabular}{c c c c}
         & &  \textbf{Pre-Test} & \textbf{Post-Test} \\
         \hline
         \hline
         \multirow{2}{*}{\textbf{`A' Cohort}} & \textbf{Men} (N = 96) & $2.84 \pm 0.04$ & $2.84 \pm 0.04$ \\
         & \textbf{Women} (N = 44) & $2.72 \pm 0.05$ & $2.77 \pm 0.05$ \\
         \multirow{2}{*}{\textbf{`B' Cohort}} & \textbf{Men} (N = 60) & $2.82 \pm 0.04$ & $2.69 \pm 0.05$ \\
         & \textbf{Women} (N = 15) & $2.8 \pm 0.1$ & $2.47 \pm 0.08$ \\
         \multirow{2}{*}{\textbf{`C, D, F' Cohort}} & \textbf{Men} (N = 51) & $2.74 \pm 0.05$ & $2.52 \pm 0.06$ \\
         & \textbf{Women} (N = 15) & $2.64 \pm 0.09$ & $2.5 \pm 0.1$ \\
    \end{tabular}
\end{table}

\section{Discussion}

As shown in Table \ref{tab:MTAresults}, the percentage of students answering each question on the calculus transfer assessment correctly increased between the beginning and the end of the semester.  Not surprisingly, student proficiency with the calculus transfer questions (question 2 and question 3) was lower than proficiency with the pure mathematics question (question 1).  This is further illustrated by the Pearson correlation coefficients shown in Table \ref{tab:correlation_post}.  The correlation between the pure mathematics question and each of the transfer questions is low, consistent with previous work indicating that mathematics transfer to physics is difficult \cite{britton2002students,nguyen2003initial,Ozimek2005,cui2006assessing,Rebello2007,new2012researching,Mikula2014}. In contrast, the transfer questions showed higher correlation with one another.  It is also worth noting that many students attempted to solve question 2 using horizontal transfer by applying an equation derived in class for a uniformly charged rod.  This contributes to lower proficiency with question 2 than question 3 and further indicates the difficulty of vertical transfer.  In other words, students may opt for horizontal transfer rather than vertical transfer since students find the former easier.  It is also possible, of course, that some students employing horizontal transfer to solve question 2 may not have read the question carefully and assumed that this problem was identical to one that they have previously solved for a uniformly charged rod.

As shown in Table \ref{tab:MTAresults}, the normalized changes in the percentage of students answering the transfer questions (question 2 and question 3) were larger for women than for men.  A larger percentage of the cohort of women did answer question 1 on the pre-test correctly, however.  This difference may be attributed to the fact that the fraction of women earning an `A' in the course (0.68) is larger than the fraction of men earning an `A' in the course (0.50).  Indeed, as shown in Figure \ref{fig:MTA_combined}, in the cohort of students earning an `A', women displayed a larger normalized gain on question 2 and question 3 than men.  In contrast, in the cohort of students earning a `B' and the cohort earning a "C, D, or F", men show a larger normalized gain for question 2 but a lower normalized gain for question 3.  

The physics self-efficacy of both men and women was lower at the end of the semester than at the beginning of the semester, consistent with what others have previously reported \cite{marshman2018female,marshman2018longitudinal,kalender2020damage}, but the magnitude of the decrease in self-efficacy was smaller for women than for men (Table \ref{tab:PSE}).  Thus, the self-efficacy of women, while consistently lower than the self-efficacy of men, was nevertheless closer to the self-efficacy of men at the end of the semester than at the beginning of the semester (Table \ref{tab:PSE}).  As with changes in the results of the calculus transfer assessment, changes in the differences between the physics self-efficacy of men and women may be attributed to a larger fraction of women earning an `A' in the course than men.  As shown in Figure \ref{fig:PSE_combined}, women earning an `A' showed an increase in their physics self-efficacy while men earning an `A' showed a decrease.  This is reflected in a decrease in the Z-statistic comparing the physics self-efficacy of men and women earning an `A' from 1.87 on the pre-test to 1.04 on the post-test.  

The physics self-efficacy of men and women in the cohort of students earning a `B' and the cohort earning a "C, D, or F" all decreased (Figure \ref{fig:PSE_combined}).  Interestingly, although women consistently exhibited a lower physics self-efficacy than that of men in all grade cohorts (Table \ref{tab:PSE}), the differences in physics self-efficacy between men and women were much smaller in the cohort of students earning a `B' and the cohort earning a "C, D, or F" than in the cohort of students earning an `A'.  Thus, the gender imbalance in physics self-efficacy increased with improved course grade.

\subsection{The Relationship Between Physics Self-Efficacy and Calculus Transfer Proficiency Is Unclear}

On the surface, these data suggest a possible inverse correlation between physics self-efficacy and calculus transfer proficiency.  For example, across all grade cohorts, women showed a larger normalized gain in calculus transfer proficiency than men, despite having a consistently lower self-efficacy.  Furthermore, although the physics self-efficacy of women in the `A' grade cohort increased, it decreased for women in all other grade cohorts.  Thus, student gender may affect the interplay between physics self-efficacy and calculus transfer proficiency, making the correlation between these attributes weaker for women than for men.  

In contrast, a different correlation between changes in calculus transfer proficiency and changes in physics self-efficacy is found for the cohort of students who answered question 3 incorrectly on the calculus transfer assessment pre-test (N = 183).  The average self-efficacy of the subset of these students who answered question 3 incorrectly on the post-test (N = 125) decreased over the semester while the average self-efficacy of the subset of these students who answered question 3 correctly on the post-test (N = 58) increased; the Z-statistic for this comparison is 2.19.  These results suggest that a positive correlation between physics self-efficacy and vertical transfer of calculus to physics (as probed by question 3 on the calculus transfer assessment) may exist.  However, the data shown in Figure \ref{fig:MTA_combined} indicate that there is no consistent correlation between the normalized change in answering question 3 correctly and either course grade or gender.

These seemingly contradictory results complicate our understanding of the relationship between physics self-efficacy and calculus transfer proficiency and thus motivate further research, including both how this relationship intersects with student identity and in the continued refinement of the instruments used to assess both calculus transfer and physics self-efficacy.

\section{Future Work}

We will build on the results presented here by incorporating into the physics self-efficacy assessment additional questions that probe mathematics self-efficacy.  We will also add additional pure math, horizontal transfer, and vertical transfer questions to the calculus transfer assessment.  We have also learned from the analysis of student responses that it would be better if the pure math questions contained integrals that more closely matched the integrals used in the solutions of the transfer problems.  It would likewise be better if we surveyed students about their gender identity and expression rather than relying upon institutional data.  Combined analysis of the results of these expanded self-efficacy and calculus transfer assessments can provide insight into the intersection of math self-efficacy, physics self-efficacy, calculus transfer proficiency, and gender.

\section*{Acknowledgments}
We would like to thank Adam Dubinsky for help in aggregating student demographic data.


\bibliography{library}

\bibliographystyle{abbrv}

\end{document}